# Giant bulk photovoltaic effect driven by interfacial symmetry breaking in $MoS_2/Ta_2NiSe_5$ heterostructures

Jianwen Ma[1#], Pengliang Leng[1#*], Lei Peng[2#], Congming Hao[3], Xianghao Meng[3], Jiaqi Liu[1], Yang Gan[4], Min Luo[5], Zifan Zhang[1], Jiaming Gu[1], Qinghang Liu[1], Lidan Duan[1], Du Xiang[4,6], Wu Shi[1,6], Peng Wang[5], Weibin Chu[2*], Xiang Yuan[3,7,8,9*], Weida Hu[5], Cheng Zhang[1,6*]

[1] State Key Laboratory of Surface Physics and Institute for Nanoelectronic Devices and Quantum Computing, Fudan University, Shanghai 200433, China

[2] Key Laboratory for Computational Physical Sciences (MOE), Institute of Computational Physical Sciences and Department of Physics, Fudan University, Shanghai 200433, China

[3] State Key Laboratory of Precision Spectroscopy, East China Normal University, Shanghai 200241, China

[4] State Key Laboratory of Integrated Chips and Systems, Frontier Institute of Chip and System, Fudan University, Shanghai, China

[5] State Key Laboratory of Infrared Physics, Shanghai Institute of Technical Physics, Chinese Academy of Sciences, Shanghai 200083, China

[6] Zhangjiang Fudan International Innovation Center, Fudan University, Shanghai 200433, China

[7] Key Laboratory of Polar Materials and Devices, Ministry of Education, School of Physics and Electronic Science, East China Normal University, Shanghai 200241, China

[8] Shanghai Center of Brain-Inspired Intelligent Materials and Devices, East China Normal University, Shanghai 200241, China

[9] Chongqing Institute of East China Normal University, Chongqing 401120, China

[#] These authors contributed equally to this work

[*] Correspondence and requests for materials should be addressed to C. Z. (E-mail: zhangcheng@fudan.edu.cn), X. Y. (xyuan@lps.ecnu.edu.cn), W. C. (wbchu@fudan.edu.cn) and P. L. (E-mail: plleng@fudan.edu.cn)

## Abstract:

Van der Waals (vdW) heterostructures offer a versatile platform for engineering unconventional bulk photovoltaic (BPV) effect through interfacial symmetry breaking. However, the coexistence of multiple photophysical mechanisms, driven by structural complexity, spontaneous charge transfer, and strong interlayer coupling, often obscures the microscopic origin of the BPV response and hinders its rational optimization. Here, we demonstrate a pronounced BPV effect localized at the overlap region of a cross-bar $MoS_2/Ta_2NiSe_5$ vdW heterostructure, where symmetry breaking induced by vertical stacking lifts the inversion center of $MoS_2$. The orthogonal device geometry enables the independent probing of intralayer and interfacial photoresponse pathways, facilitating clear separation of competing mechanisms. Spontaneous interfacial charge transfer between $MoS_2$ and $Ta_2NiSe_5$ further establishes a strong interlayer electronic coupling. By modulating the interlayer potential landscape through gate voltage and vertical electric fields, we achieve an optimized zero-bias photocurrent density of 247 $A/cm^2$ and a BPV coefficient of 0.99 $V^{-1}$. Supported by theoretical modelling, our results illustrate how minimalist device geometry can transform complex heterostructures into experimentally tractable platforms. This strategy paves the way for analyzing and optimizing interface-driven BPV effect, with implications for self-powered optoelectronics, broadband photodetection, and energy-harvesting nanodevices.

## Introduction:

The bulk photovoltaic (BPV) effect, which manifests as spontaneous photocurrent generation under zero external bias, fundamentally arises from the shift current mechanism inherent to non-centrosymmetric crystal structures[1–4]. Owing to its intrinsic nonequilibrium nature, BPV effect has attracted sustained interest as a potential route to surpass the Shockley-Queisser efficiency limit[5], with implications for next-generation energy harvesting and optoelectronics[6,7], establishing it as a forefront topic at optoelectronic study[8–10]. Despite this conceptual generality, experimentally observable BPV effect has so far been largely restricted to a limited set of material classes, including ferroelectric oxides[10–13], polar semiconductors[14–19], and certain topological materials[20–23] or a few low-dimensional vdW systems[24–26].

Recent advances have highlighted interfacial engineering as a promising route to induce BPV responses beyond the constraints imposed by intrinsic crystal symmetry[27–29]. In vdW heterostructures, inversion symmetry can be locally broken through twist-angle control, strain engineering, or the formation of polar interfaces[30–33], enabling BPV effect to emerge even in materials that are centrosymmetric in isolation. While this strategy significantly expands the accessible material space and offers new opportunities for junction-free and gate-tunable optoelectronic devices, it also introduces substantial complexity. Stacking two various layers inevitably leads to strong interlayer coupling, including spontaneous charge transfer and interlayer electronic hybridization[28], which can profoundly modify carrier generation, separation and recombination pathways. As a result, multiple photoresponse mechanisms often coexist and overlap spatially within a single heterostructure, making it challenging to disentangle interfacial and intralayer contributions or to establish clear optimization principles for interface-driven BPV effect.

Here we address this challenge by combining interfacial symmetry breaking with a cross-bar device geometry that enables the observation of the giant BPV effect as well as independent analysis of multiple photovoltaic processes. Leveraging the strong in-plane anisotropy inherent to $Ta_2NiSe_5$ by vertically stacking onto $MoS_2$, the cross-bar architecture of $MoS_2/Ta_2NiSe_5$ heterostructure generates a pronounced interfacial BPV effect resulting from symmetry breaking at the overlap region (Fig. 1a). The significant lattice mismatch and interfacial interactions between $MoS_2$ and $Ta_2NiSe_5$ reduces $MoS_2$'s intrinsic $D_{3h}$ symmetry to $C_s$ symmetry at the interface. This symmetry reduction induces a net in-plane polarization (red arrow in Fig. 1b) parallel to the surviving mirror plane ($M_{ab}$), enabling unconventional photoresponse behaviors via symmetry-forbidden carrier injection.[34] Within this framework, the BPV response exhibits a zero-bias photocurrent density of 247 A/cm$^2$ and a BPV coefficient of 0.99 V$^{-1}$. By explicitly separating the roles of interfacial symmetry breaking and charge-transfer-induced fields, this work establishes a general experimental strategy for analyzing and optimizing BPV effect in complex vdW heterostructures.

## Results and Discussion

The monolayer 2H-$MoS_2$ exhibits a hexagonal lattice (point group $D_{3h}$), characterized by three-fold rotational symmetry ($C_3$) and mirror plane ($M_{ab}$) parallel to the $a$-axis direction[35]. In contrast, $Ta_2NiSe_5$ adopts an orthorhombic structure (space group $C_{mcm}$) with two-fold rotational symmetry ($C_2$) along its quasi-1D chain direction ($a$-axis) and mirror planes parallel to the chains direction[36] (Supplementary Figure 1). $Ta_2NiSe_5$ conceives a small direct gap[37], resulting in a high density of states near the Fermi level that endows the material with quasi-metallic conductivity[38,39]. Crucially, the strong in-plane anisotropy was verified by resistivity measurements[36], optical characterizations[40], tensile tests[41], and further confirmed by angle-resolved Raman measurement (Fig. 1c), a direct consequence of its quasi-one-dimensional chain-like crystal structure aligned along the $a$ axis[42] (Supplementary Figure 2). This synergistic combination of narrow-gap-enhanced carrier density and strong anisotropy establishes $Ta_2NiSe_5$ as an ideal platform for engineering symmetry-broken interfacial phenomena[43,44].To assess crystalline quality and interfacial integrity of $MoS_2$/$Ta_2NiSe_5$ heterostructure, we performed spatially resolved Raman spectroscopy across pristine and overlapped regions (Fig. 1d). The isolated $MoS_2$ flake exhibits characteristic phonon modes at 385 $cm^{-1}$ (in-plane $E_{2g}^{1}$) and 407$cm^{-1}$ (out-of-plane $A_{1g}$)[45]. Similarly, $Ta_2NiSe_5$ displays signature peaks at 95.5 $cm^{-1}$ ($A_g^1$) and 121.1 $cm^{-1}$ ($A_g^2$), consistent with single-crystal standards[46]. Critically, the heterointerface spectrum retains all fundamental modes of both materials, confirming a structurally coherent vdW heterostructure free of lattice distortion (Supplementary Figure 3). Notably, the attenuation of Raman intensity at the overlap region arises from interfacial charge-transfer-induced lattice softening, which reduces Raman scattering cross-sections through phonon damping[47].

To reveal the underlying physical mechanism of BPV effect in this system, we constructed a $MoS_2$/$Ta_2NiSe_5$ heterostructure model and performed comprehensive electronic structure and nonlinear optical response analyses. The calculated flat band structure of the heterostructure (Fig. 1e) reveals a typical type-I band alignment. Notably, the states near the Fermi level are predominantly contributed by the $Ta_2NiSe_5$ bands, indicating that the low-energy excitations and transport properties are primarily governed by the $Ta_2NiSe_5$ layer. A pronounced energy dispersion difference is observed between the Γ−X and Γ−Y directions, directly confirming the in-plane anisotropy of the system. The interfacial stacking between $MoS_2$ and $Ta_2NiSe_5$ breaks inversion symmetry and allows finite shift-current conductivity components. The anisotropic $Ta_2NiSe_5$ derived band dispersion further modulates the magnitude and polarization dependence of the allowed response through direction dependent optical transition matrix elements. The calculated shift current conductivity tensor $\sigma_{acc}$ and $\sigma_{aaa}$ components (Fig. 1f), corresponding to $a$-axis photocurrent generated by $c$- and $a$-polarized light respectively, characterize the intrinsic BPV response of the heterostructure. The calculated spectra for $\sigma_{\mathrm{acc}}$ and $\sigma_{\mathrm{aaa}}$ exhibit significant peaks across a broad energy range. The shift current conductivity is significantly observable at the characteristic energy of experimental 633-nm excitation (photon energy ~1.96 eV), regardless of whether the excitation is applied along the *x*-direction or the *y*-direction of the device. To gain microscopic insight into the origin of these responses, we further examine band-resolved shift current contribution maps, which highlight the critical role of specific interband transitions in generating the observed photocurrent (Supplementary Figure 4).

## Symmetry-broken induced BPV effect in $Ta_2NiSe_5$

A key challenge in vdW heterostructures is the entanglement of interlayer charge transfer, built-in fields, and anisotropic intralayer transport, which obscures the origin and optimization of BPV effect. To address this, we design a cross-bar $MoS_2/Ta_2NiSe_5$ heterojunction (Supplementary Figure 5), where the orthogonal overlap geometry spatially separates interfacial photocurrent generation from electrode-contact effects and enables independent access to both interlayer and intralayer photophysical channels.

Upon forming the van der Waals heterostructure, the *a* axis of $Ta_2NiSe_5$ was stacked roughly perpendicular to the zigzag direction of $MoS_2$, as schematized in Fig. 2a, which is determined by second harmonic generation (SHG) measurements (Supplementary Figure 6). Under illumination, the BPV response emerges at the $MoS_2/Ta_2NiSe_5$ overlap interface, where interfacial charge transfer and orbital hybridization lower the symmetry and enable nonzero in-plane shift-current tensor components along *a* axis. In a cross-bar $MoS_2/Ta_2NiSe_5$ heterojunction, BPV effect measurement was performed directly along the two terminals of $Ta_2NiSe_5$, which is parallel with in-plane polarization direction. Current-voltage (*I-V*) characteristics under various power illumination were first measured by applying bias directly across two $Ta_2NiSe_5$ terminals (Fig. 2b) in Device A. The dark current exhibits linear Ohmic behavior (slope $R$ = 28.2 kΩ), confirming barrier-free contacts and metallic property[48]. Under 633-nm laser illumination focused at the overlap center, we observe a substantial open-circuit voltage $V_{OC}$ = 11.20 mV and short-circuit current $I_{SC}$ = 408.17 nA at $P$ = 50 μW, unambiguously evidencing the emergence of giant zero-bias photocurrent generation. Moreover, polarization-resolved photovoltage measurements further reveal a typical two-fold symmetry that coincides with the $Ta_2NiSe_5$ *a*-axis and the interface polar axis (Supplementary Figure 7).

To spatially resolve the interfacial BPV effect, we performed scanning photocurrent microscope (SPCM) measurements across the $MoS_2/Ta_2NiSe_5$ heterostructure (Fig. 2c). A typical optical microscope image of the fabricated $MoS_2/Ta_2NiSe_5$ heterojunction Device B is shown in the inset of Fig. 2c. The regions of pristine $MoS_2$ and $Ta_2NiSe_5$ are delineated by dashed blue and purple contours, respectively. The 2D photovoltage map reveals pronounced zero-bias photovoltage with uniform polarity, confined to the $MoS_2/Ta_2NiSe_5$ overlap region. The spatial photovoltage mapping confirms that the zero-bias photoresponse is localized within the $MoS_2/Ta_2NiSe_5$ overlap region, further excluding contact-dominated photovoltaic contributions. The line profile of photovoltage (Fig. 2d) demonstrates abrupt $V_{OC}$ transitions at overlap boundaries and shows no correlation with electrode-contact edges, ruling out conventional photovoltaic mechanisms from electrodes. Additionally, we compare the photovoltage mappings of two $Ta_2NiSe_5$ devices with different thicknesses, another thin $MoS_2$ device and a misaligned-stacking $MoS_2/Ta_2NiSe_5$ heterostructure device; in all cases, the photovoltage signals are predominantly localized at the electrode contact and edge regions (Supplementary Figures 8-11). In this heterostructure, bottom $MoS_2$ serves as the primary optical absorption layer, while the $MoS_2/Ta_2NiSe_5$ interface generates net directional photocurrent. We also find that the photovoltage is not detectable under 1.5 μm infrared illumination (Supplementary Figure 12), consistent with inefficient photocarrier generation in few-layer $MoS_2$ at a photon energy of ~0.8 eV, below its optical gap of ~1.2-1.5 eV[49]. Although the calculated shift-current spectrum shows finite response near this energy, the weak optical absorption under this excitation limits the experimentally measurable photovoltage. Combined with symmetry analysis and material exclusion measurements in control devices, our results point to a spontaneous in-plane photovoltaic mechanism arising from the $MoS_2/Ta_2NiSe_5$ interface.

The laser power dependence of the $V_{OC}$ in our cross-bar heterostructure exhibits a distinct crossover from linear to sublinear (square-root) scaling at approximately $P \approx$ 2 μW (Fig. 2e). This characteristic transition cannot be attributed to conventional mechanisms such as Schottky barrier photovoltaics near contacts, which typically yield purely linear power dependence, but instead points to the shift current mechanism inherent to non-centrosymmetric systems[50]. The observed crossover can be understood as a saturation effect of photoexcited carriers at higher optical intensity, as predicted in the shift-current model and observed experimentally[28,29]. On-off cycling stability of $V_{OC}$ was systematically investigated under increasing laser power (from 1.5 to 35 μW, Fig. 2f). All cycles maintain identical $V_{OC}$ rise/decay profiles with zero amplitude degradation. This exceptional stability originates from the symmetry-broken interfacial polarization mechanism, which is topologically protected by lattice mismatch, preventing screening-induced weakening.

## Interlayer charge transfer in $MoS_2/Ta_2NiSe_5$ interface

The narrow-gap band structure of $Ta_2NiSe_5$ generates a high density of states (DOS) at Fermi level ($E_F$), facilitating pronounced spontaneous interlayer charge transfer (Fig. 3a) when stacking vertically onto $MoS_2$, forming a type-I band bending[37,51] (Fig. 3b) at the interface, where the Fermi level difference drives electrons transfer from $MoS_2$ to $Ta_2NiSe_5$. Consequently, a vertical built-in electric field is established between $MoS_2$ and $Ta_2NiSe_5$. To probe this effect, we applied a vertical electrical field ($V_z$) perpendicular to the interface by connecting $MoS_2$ and $Ta_2NiSe_5$ terminals and carried out the electrical *I-V* measurement. As shown in Fig. 3c, the device exhibits strong rectification, originating from the interfacial built-in field. While this field does not directly drive long-range electron-hole separation, it modulates the interlayer hybridization and local carrier dynamics, thereby influencing recombination pathways and photoresponse efficiency when the overlap region is illuminated. Thus, this interfacial charge transfer process serves as a critical enabler of the rich and tunable photoresponse behaviors observed in such heterostructure devices[34,52].

Gate-dependent modulation of the vertical barrier was achieved by applying back-gate voltages ($V_{BG}$) through the $SiO_2$/Si substrate to control carrier concentration in the bottom $MoS_2$ layer. Figure 3d presents the evolution of *I-V* characteristics when connecting the $MoS_2$ and $Ta_2NiSe_5$ terminals as $V_{BG}$ increases from -10 V to 15 V, as shown in the inset of Fig. 3c. Under forward bias conditions, the saturation current exhibits significant enhancement with increasingly positive $V_{BG}$, whereas in the reverse bias ohmic regime, variations in current slope directly reflect gate-tunable series resistance. As a result, we quantitatively extracted the gate-modulated barrier height from saturation current values, revealing a systematic reduction from 0.53 eV at $V_{BG}$ = -10 V to 0.44 eV at $V_{BG}$=15 V (Fig. 3e), demonstrating effective electrostatic control over interfacial band alignment.

The photovoltaic voltage from vertical built-in field at the overlap region of $MoS_2$/$Ta_2NiSe_5$ heterostructure was systematically characterized by measuring $MoS_2$ and $Ta_2NiSe_5$ terminal. A spatially uniform photovoltage distribution with preserved polarity is observed throughout the junction region (Fig. 3f). By the way, although Fig. 2c and Fig. 3f both show signals localized mainly in the $MoS_2$/$Ta_2NiSe_5$ overlap region, they were obtained under different electrical configurations and probe different photovoltaic responses (Supplementary Figure 13). The apparent spatial similarity between the two maps arises because both effects are generated most efficiently in the overlap region where the $MoS_2$/$Ta_2NiSe_5$ heterointerface is formed. However, the former detects the in-plane BPV response along the $Ta_2NiSe_5$ channel, while the latter probes the vertical junction photovoltaic response associated with interfacial charge transfer and the resulting built-in electric field. The line profile in Fig. 3g demonstrates that the photovoltage magnitude peaks at the junction center while maintaining consistent polarity. Figure 3h presents the output characteristics under 633-nm laser illumination across laser power ranging from 0 to 9.32 μW. With increasing optical power, the *I-V* curves systematically shift parallelly upward while maintaining their fundamental shape, indicating robust photoresponse sensitivity and operational stability.

## Lateral carrier concentration gradient in $MoS_2$

The cross-bar geometry also allows the detection of in-plane electronic landscapes in the constituent layers. Specifically, interfacial charge transfer from $MoS_2$ to $Ta_2NiSe_5$ induces laterally modulated carrier concentration in the bottom $MoS_2$ layer. It gives rise to an N-$N^-$-N junction that spans the non-overlapping, overlapping, and opposite non-overlapping regions along the $MoS_2$, as shown in Figure 4a. In the overlap zone, electron depletion creates a local minimum in carrier density, flanked by electron-rich regions on either side, forming a high-low-high electron concentration profile visualized by a color gradient of the underlying layer. Fig. 4b illustrates a schematic band bending in the $MoS_2$ layer induced by the gradient of electron concentration. This spatially graded doping landscape is directly probed via two-terminal electric measurements along the $MoS_2$ channel (Fig. 4c).

The electrical *I-V* measurement was carried out by connecting two terminals of bottom $MoS_2$, which displays quasi-ohmic behavior at low voltage, transitioning to progressive current saturation under elevated forward and reverse voltages (Fig. 4c). This saturation stems from charge-accumulation-induced barrier at the N/N$^-$ interfaces: under forward bias ($V$ > 0.5 V), excess electron accumulation in the N$^-$ region, raising the first junction barrier and suppressing further injection; under reverse bias, an analogous potential barrier forms at the opposite N/N$^-$ boundary, yielding symmetric saturation currents. Back gate modulation experiments (Fig. 4d-e) confirm N-N$^-$-N distribution: $V_{BG}$ (-10 to 15 V) systematically enhances saturation currents in both polarities, corresponding to the barrier height from 0.51 eV to 0.41 eV calculated from the saturation currents under positive bias voltages. The calculation details are provided in Supplementary Note 11. Spatially resolved photovoltage mapping (Fig. 4f) reveals that illumination-generated electron-hole pairs are bidirectionally separated by opposing in-plane lateral built-in fields at the two N/N$^-$ ends. This results in a polarity reversal of photovoltage across the depletion zone, corresponding to the drift-driven carrier separation under lateral band bending. The photoresponse of junction formed by the N/N$^-$/N configuration can also be modulated by the incident optical power under 633-nm illumination, as shown in Figure 4h (Supplementary Figure 14). The slight asymmetry in the $I$-$V$ curves and the unequal photovoltage magnitudes at the two overlap edges may arise from the non-ideal geometry of the actual $MoS_2$/$Ta_2NiSe_5$ overlap region. The two N/N$^-$ junction boundaries in $MoS_2$ are not perfectly equivalent due to the unequal contact-edge lengths, which may lead to asymmetric local electric-field distributions, series resistances, and carrier-collection efficiencies.

**Giant BPV enhancement via synergistic optoelectronic modulation**

Beyond conventional symmetry-broken BPV effect at overlapping region in $Ta_2NiSe_5$ when vertically stacking onto $MoS_2$, our platform uniquely integrates two synergistic optoelectronic control dimensions: a charge-transfer-induced vertical built-in electric field at interface superimposed on the lateral electric field arising from the charge concentration gradient in bottom $MoS_2$. This dual-channel architecture enables unprecedented optoelectronic enhancement by applying a back-gate voltage ($V_{BG}$) at silicon substrate and a vertical voltage ($V_z$) from bottom $MoS_2$ to top $Ta_2NiSe_5$ terminal independently (Fig. 5a). By co-optimizing these parameters ($V_{BG}$ = -10 V, $V_z$ = 1 V), we achieve a remarkable enhancement in open-circuit voltage ($V_{OC}$) by over an order compared to $V_{OC}^0$ ($V_{BG}$ = $V_z$ = 0 V) (Supplementary Figure 15). We tentatively attribute the pronounced BPV enhancement to the combined electrostatic modulation by $V_{BG}$ and $V_z$. $V_{BG}$ mainly tunes the carrier density and electrostatic environment of the bottom $MoS_2$ layer, thereby modifying the interfacial charge transfer, screening, and the non-centrosymmetric potential landscape of the $MoS_2$/$Ta_2NiSe_5$ overlap region. This modulation changes the conditions under which the in-plane BPV response is generated at the symmetry-broken interface. By contrast, positive $V_z$ enhances the intrinsic charge-transfer-induced built-in field, raising the barrier for electron transfer from $MoS_2$ to $Ta_2NiSe_5$. This allows the laterally generated BPV signal to be collected more efficiently. The present qualitative analysis does not explicitly account for possible bias-induced changes in the intrinsic BPV generation process; evaluating these effects will require further theoretical and experimental investigation.

To quantitatively characterize the observed photoresponse within the framework of BPV effect in our $MoS_2/Ta_2NiSe_5$ heterostructure, we derive the BPV coefficient $\beta$ from $j_q = \beta_{qrs} e_r e_s^* I_0$, where $j_q$ is the photocurrent density, $e_r$ and $e_s^*$ are the light polarization unit vectors, and $I_0$ is the incident laser power density. The photocurrent density ($j_{sc}$) is calculated by normalizing to the cross-sectional area through which photocurrent flows in the top $Ta_2NiSe_5$ layer[53,54]. Here, the diameter of the 633-nm incident laser is 1.6 μm. A comparative analysis of the BPV performance in the $MoS_2/Ta_2NiSe_5$ heterostructure is presented alongside three distinct material categories, where respective $\beta$ coefficients and photocurrent density are denoted in Figures 5c-d by half-filled, open, and solid symbols, respectively. The first class is classical non-centrosymmetric materials[12,13,55–59], exemplified by $BaTiO_3$ (BTO)[57], which exhibit spontaneous polarization. However, the inherent BPV coefficients in this class are generally constrained to lower magnitudes due to fundamental limitations associated with their polarization mechanisms and band structures. The second class is materials with locally engineered symmetry breaking[9,32,60–62], typified by strain-gradient-engineered $MoS_2$ nanosheets[9]. This approach deliberately utilizes controlled strain or flexo configurations to introduce strain gradients that break local inversion symmetry, thereby satisfying the fundamental prerequisite for generating a BPV current and promoting the directional separation of photogenerated carriers. While the synergistic interplay between intrinsic material properties (*e.g.*, symmetry characteristics, light absorption capacity) and external modulation (*e.g.*, strain engineering, device architecture) can yield excellent BPV performance, the scalability and practical application potential of this strategy are significantly hampered by the inherently limited active area over which the symmetry is effectively broken. The last class is the vdW heterostructures[27–29], represented by a combination such as $WSe_2$/black phosphorus (BP)[27]. These structures not only demonstrate superior optoelectronic characteristics but also offer the advantage that their effective area is primarily determined by overlapping areas in device fabrication and stacking processes. Crucially, by judiciously selecting constituent layers with specific symmetries, a robust interfacial BPV effect can be realized across extensive interfacial areas, enabling scalable device integration.

Our $MoS_2/Ta_2NiSe_5$ heterostructure belongs to this third category. Its performance is competitive with established vdW counterparts. The strategic selection of $Ta_2NiSe_5$, characterized by its pronounced in-plane anisotropy and narrow bandgap, combined with $MoS_2$ stacking, effectively breaks the interfacial symmetry. Furthermore, the narrow-bandgap nature of this vertical structure engenders a high DOS near the Fermi level. This condition facilitates exceptionally efficient interfacial charge transfer, establishing a substantial built-in electric field at the interface, which further amplifies the BPV response. Significantly, within this straightforward cross-bar device configuration, the BPV performance can be substantially enhanced through the additional modulation of carrier concentration within the underlying $MoS_2$ layer and the precise tuning of the vertical barrier height. Consequently, this work underscores the critical importance of both meticulous material selection and thoughtful device architecture design for achieving superior BPV effect in vdW heterostructures.

In conclusion, we demonstrate a tunable giant BPV effect in a cross-bar shaped $MoS_2/Ta_2NiSe_5$ heterostructure, achieving an unprecedented photocurrent density of 247 A/cm$^2$ and BPV coefficient of 0.99 V$^{-1}$ that fundamentally originates from interfacial symmetry breaking. The choice of $Ta_2NiSe_5$ with strong in-plane anisotropy and narrow-gap band structure renders spontaneous interlayer charge transfer from bottom $MoS_2$ to top $Ta_2NiSe_5$, which facilitates a net in-plane polarization within $MoS_2$ /$Ta_2NiSe_5$ interface. The design of a cross-bar shape heterojunction provides access to explicitly figure out the roles of interfacial symmetry breaking and charge-transfer-induced fields. Consequently, the dual-channel electrostatic tunability of this architecture enables unprecedented performance modulation: negative back-gate voltages ($V_{BG}$) reconfigure the carrier concentration gradient, while positive vertical voltage ($V_z$) reshapes the barrier height of vertical built-in field, collectively achieving over ten times enhancement in zero-bias open-circuit photovoltage. Our findings demonstrate interfacial engineering as a powerful strategy for optimizing the BPV effect. This approach not only provides a clear pathway toward optimizing interfacial BPV systems but also opens avenues for designing reconfigurable nano-optoelectronic devices.

# Methods

## Device fabrication

The $MoS_2/Ta_2NiSe_5$ heterostructure devices were fabricated using mechanically exfoliated materials through a Polycarbonate-assisted transfer method. Firstly, $MoS_2$ or $Ta_2NiSe_5$ flakes were mechanically exfoliated from single crystals. Secondly, the bottom $MoS_2$ layer was transferred onto the pre-fabricated $SiO_2/Si$ substrate with Au electrodes under an optical microscope. Then, the top $Ta_2NiSe_5$ layer was transferred onto the bottom $MoS_2$ using the same method to form a crossing heterostructure. The bottom Cr/Au (5 nm/10 nm) electrodes were fabricated using standard electron-beam lithography (EBL), followed by electron-beam evaporation for metal deposition and a subsequent lift-off process.

## SPCM and electrical measurements

The SPCM measurement was carried out by a home-built scanning microscope. A He-Ne laser (633 nm) was chopped by a mechanical chopper at 380 Hz, and then focused onto the device to induce photovoltage generation at room temperature. The laser beam had a spot diameter of about 1.6 µm, and the motion of the focal spot was controlled by voltage-driven galvo scanners. The generated photovoltage was collected by a lock-in amplifier (Stanford SR865) at the chopped frequency. Meanwhile, the reflected laser light was monitored by a photodetector to determine the laser spot position on the sample surface.

## Single crystal growth

Single crystals of $Ta_2NiSe_5$ were synthesized via a chemical vapor transport (CVT) method. Elemental powders of tantalum, nickel, and selenium were mixed in a stoichiometric ratio of approximately 1.3:0.2:1.4 and sealed in an evacuated quartz ampoule (~$1 \times 10^{-3}$ Pa) with a small amount of iodine serving as the transport agent. The sealed ampoule was subjected to a temperature gradient of 950 °C/850 °C for one week. Upon completion of the transport process, needle-like $Ta_2NiSe_5$ single crystals were obtained at the cooler end of the tube.

## Theoretical calculations

Theoretical investigations were performed using Density Functional Theory as implemented in the Atomic orbital Based Ab-initio Computation at UStc (ABACUS) package[67]. The exchange-correlation interactions were treated with the Perdew-Burke-Ernzerhof (PBE) functional[68]. The numerical atomic orbital (NAO) basis set and optimized norm-conserving Vanderbilt (ONCV) pseudopotentials are used[69–71]. An energy cutoff of 120 Ry and a $\Gamma$-centered $2\times3\times2$ k-point grid are used for self-consistent and band structure calculations.

The shift current conductivities, representing the primary intrinsic contribution to the BPV effect, were calculated using the Py-ATB package based on the independent-particle approximation[72]. The shift current tensor $\sigma^{\mathrm{abc}}$ is expressed as:

$$\sigma^{\mathrm{abc}}(0;\omega,-\omega)=\frac{\pi e^3}{\hbar^2}\int\frac{\mathrm{d}k}{8\pi^3}\sum_{\mathrm{n,m}}f_{\mathrm{nm}}R_{\mathrm{nm}}^{\mathrm{a,bc}}r_{\mathrm{nm}}^{\mathrm{b}}r_{\mathrm{mn}}^{\mathrm{c}}\delta(\omega_{\mathrm{mn}}-\omega)$$

where $r_{\mathrm{nm}}^{\mathrm{b}}$ is the inter-band dipole matrix element and $R_{\mathrm{nm}}^{\mathrm{a,bc}}$ is the generalized derivative of the dipole matrix related to the quantum geometry of the Bloch wave functions. The tight-binding Hamiltonian used for the response calculations was directly constructed from the ABACUS self-consistent results.

## Data availability

The Source Data underlying the figures of this study are available with the paper. All raw data generated during the current study are available from the corresponding authors upon request.

## Acknowledgements

Part of the sample fabrication was performed at Fudan Nanofabrication Laboratory.

## Funding

This work was supported by the National Key R&D Program of China (Grant No. 2022YFA1405700, 2023YFA1407500), the National Natural Science Foundation of China (Grant No. 92365104, 12174069), the Shanghai Pilot Program for Basic Research-Fudan University 21TQ1400100 (25TQ001), Innovation Project for Integration of Science and Education from Shanghai Institute of Technical Physics, CAS (SITPKJRH-2025-02), and Shanghai QiYuan Innovation Foundation. P.L. was supported by the National Natural Science Foundation of China (Grant No. 12504188), the Natural Science Foundation of Shanghai (Grant No. 25ZR1402017), and the China Postdoctoral Science Foundation (Grant No. 2025M783367). X.Y. was also supported by the National Natural Science Foundation of China (Grant No. U24A2012), the Scientific Research Innovation Capability Support Project for Young Faculty (Grant No. ZYGXQNJSKYCXNLZCXM-M11), Shanghai Pilot Program for Basic Research (Grant No. TQ20240203), Shanghai Rising-Star Program (Grant No. 24QA2702200), Shuguang Program (Grant No. 24SG29). W.C. was supported by Shanghai Pilot Program for Basic Research-Fudan University 21TQ1400100 (22TQ017), and the National Natural Science Foundation of China (Grant No. 12188101, 12274081).

## Author Contributions

C.Z. conceived the ideas and supervised the overall research. X.M., C.H., and X.Y. synthesized the crystals and performed optical characterization. J.M. fabricated and characterized the device with the help of P.L., J.L., Y.G., Z.Z., J.G., Q.L, L.D., D.X. and W.S.. L.P. and W.C. performed theoretical calculations. J.M., P.L., M.L., X.Y., P.W., W.H. and C.Z. analysis the data. J.M., P.L., L.P., W.C. and C.Z. wrote the paper with assistance from all other co-authors.

## Competing interests

The authors declare no competing interests.

**Figure 1 | Schematic cross-bar $MoS_2/Ta_2NiSe_5$ heterostructure device and interfacial photovoltaic mechanism. a,** Device architecture with $MoS_2$ (bottom) and $Ta_2NiSe_5$ (top) forming overlapping regions. Illumination by a 633-nm-wavelength laser (spot indicated) triggers interfacial bulk photovoltaic (BPV) effect. Electrical measurements connect through bottom electrodes. This schematic was created using Blender. **b,** Atomic structure of $MoS_2$ (bottom) and $Ta_2NiSe_5$ (top). $MoS_2$ exhibits in-plane threefold rotational symmetry ($C_3$) and mirror plane $M_{ab}$, while $Ta_2NiSe_5$ possesses a twofold rotational axis parallel to its quasi-1D chain direction (*a* axis). The symmetry at the overlapping region reduces to $C_s$ with mirror plane $M_{ab}$, generating in-plane polarization (red arrow). The atoms are labeled. **c,** Angle-resolved Raman measurements in $Ta_2NiSe_5$ nanostructure show strong in-plane anisotropy. Blue dots, experimental data; red dashed line, fitted data. **d,** Raman mapping of characteristic peaks: $MoS_2$ region ($E_{2g}^1$ at 385 cm$^{-1}$, $A_{1g}$ at 407 cm$^{-1}$, deep blue), $Ta_2NiSe_5$ region ($A_g^1$ at 95.5 cm$^{-1}$, $A_g^2$ at 121.13 cm$^{-1}$ light green), and overlap zone (orange). Dashed lines highlight peak positions. **e,** Calculated flat band of the heterostructure. The color scale indicates the orbital contribution from $MoS_2$ (blue) and $Ta_2NiSe_5$ (red), highlighting the type-I alignment and anisotropic dispersion along Γ-X and Γ-Y directions. **f,** Calculated shift current conductivity spectra for $\sigma_{\text{aaa}}$ and $\sigma_{\text{acc}}$ components, characterizing the intrinsic BPV response.

**Figure 2 | Interfacial BPV effect in $MoS_2/Ta_2NiSe_5$ heterostructures. a,** The measurement configuration of BPV effect, high potential terminal is labeled by "+", low potential terminal is labeled by "-". Schematic created using Blender. **b,** *I-V* curves under varied laser power densities (*P*), obtained via DC measurements in Device A. **c,** Spatially resolved open-circuit voltage ($V_{OC}$) mapping shows enhanced BPV signal exclusively at overlap regions. Inset: Optical micrograph of a corresponding cross-bar device. Scale bar: 10 μm. **d,** Line profile along $Ta_2NiSe_5$ layer, confirming localized BPV response at overlap zones. **e,** Open-circuit voltage ($V_{OC}$) under increasing *P* (solid line: power-law fit). **f,** $V_{OC}$ on-off cycling stability under increasing power (1.5 to 35 μW). Data in **c-f** are representative results from Device B characterized by lock-in AC techniques to ensure high signal-to-noise ratio.

**Figure 3 | Interfacial charge transfer in $MoS_2/Ta_2NiSe_5$ heterostructures. a,** Cross-sectional schematic of vertical electron transfer from bottom $MoS_2$ to top $Ta_2NiSe_5$, where interfacial charge redistribution creates a built-in electric field. **b,** Type-I band bending establishes a built-in field. Schematic band diagram showing $E_C$ (conduction band), $E_V$ (valence band), and $E_F$ (Fermi level). Red and blue circles denote holes and electrons, respectively, and the arrow indicates the charge transfer direction. **c,** Current-voltage (*I-V*) characteristics, colored regions indicate the areas between the *I-V* curve and the horizontal axis (cyan: positive; purple: negative). Inset: the measurement configuration, high potential terminal is labeled by "+", low potential terminal is labeled by "-", the inset was prepared using Blender. **d,** Back-gate voltage ($V_{BG}$) modulated *I-V* curves of vertical built-in electric field from -10 V to 15 V. **e,** Quantitative relationship between extracted barrier height $E_{\text{barrier}}$ and $V_{BG}$. Barrier height versus gate voltage, presented as a dot-line plot. **f,** Spatially resolved photovoltage scan from $MoS_2$ to $Ta_2NiSe_5$ terminal. **g,** Line profile along $MoS_2$ layer. **h,** Complete photovoltaic *I-V* characteristics under various laser powers.

**Figure 4 | Gate-tunable transport and photoresponse in lateral $MoS_2$ N-$N^-$-N junctions. a**, Spatial distribution of electron concentration in bottom $MoS_2$, where the central light blue region corresponds to a relatively low electron density and the blue regions on two sides indicate areas of higher electron concentration. **b,** Schematic band diagram illustrating charge-transfer-induced electron depletion in the overlapping region. The arrow denotes the charge transfer direction. Red and blue circles correspond to holes and electrons, respectively. **c,** Two-terminal *I-V* characteristics of the $MoS_2$ channel confirming N-$N^-$-N junction formation. The inset is the measurement configuration ,where the high potential terminal is labeled by "+", the low potential terminal is labeled by "-", and it was produced using Blender. The colored regions highlight the areas of the *I-V* curve under positive and negative bias voltages. **d,** Gate-voltage ($V_{BG}$ = -10 to 15 V) modulation of saturation currents in both polarities. **e,** Electrostatic control of barrier height calculated from the saturation currents under positive and negative bias voltages. **f,** Scanning photovoltage mapping showing polarity reversal across the $MoS_2$. **g,** Line profile of (f) with symmetric $\pm\Delta V$ peaks indicating bidirectional carrier separation. **h,** 633-nm power-dependent photocurrent demonstrating lateral band-gradient separation.

**Figure 5 | Back-gate voltage ($V_{BG}$) and vertical voltage ($V_z$) co-modulation of BPV effect. a,** The measurement configuration of applied back gate voltage ($V_{BG}$) and vertical voltage ($V_z$). The $V_z$ was applied from bottom $MoS_2$ terminal to top $Ta_2NiSe_5$ terminal. **b,** Back-gate ($V_{BG}$ = -10 to 15 V) dependence of open-circuit voltage ($V_{OC}$) at fixed bias ($V_z$ = 0 to 1 V), showing giant enhancement at negative $V_{BG}$ and positive $V_z$. **c,** Experimental BPV coefficients $\beta$ for various materials. Data for non-centrosymmetric materials ($BaTiO_3$ (BTO)[57], $Pb(Zr_xTi_{1-x})O_3$ (PZT)[63], $BiFeO_3$ (BFO)[12], $KBiFe_2O_5$ (KBFO)[55], Sb-doped ZnO (ZnO:Sb)[58], $SnP_2Se_6$[16], $SnP_2S_6$[17], $AgBiP_2Se_6$[18], $CuInP_2Se_6$[19]), bulk materials ($WS_2$ nanotube (NTs)[60], strain-gradient-engineered $MoS_2$ (sg)[9], 3R-$MoS_2$[33,62]), and heterostructure ($WSe_2$/Black phosphorous(BP)[27], $MoS_2$/BP[28], $In_2Se_3$/Graphene(Gr)[64]) are shown as half-filled, open and empty symbols, respectively. Data for BTO and ZnO are $\beta_{31}$, for BFO is $\beta_{22}$, and for others are effective values. **d,** Photocurrent density $j_{SC}$ versus laser power density. Data for non-centrosymmetric materials (methylammonium lead iodide ($MaPbI_3$)[59], BFO[12], $Bi_2FeCrO_6$ (BFCO)[13], Bi-Mn-O composite thin film[65], KBFO[55], BTO[57], ZnO:Sb[58], PZT[63], $SnP_2Se_6$[16], $SnP_2S_6$[17], $AgBiP_2Se_6$[18], $CuInP_2Se_6$[19]), bulk materials ($WS_2$ NTs[60], $MoS_2$ sg[9], 3R-$MoS_2$[33,62]), and heterostructure ($WSe_2$/BP[27], $MoS_2$/BP[28], $In_2Se_3$/Gr[64]) are shown as half-filled, open and empty symbols, respectively. The solid orange and pink lines correspond to PZT and BTO[66], respectively. The $MoS_2/Ta_2NiSe_5$ data in **c** and **d** are highlighted by light-red-shaded red balls, which were obtained from Device A at incident laser powers of 5, 10, 30 and 50 μW, without external bias and gate modulations.

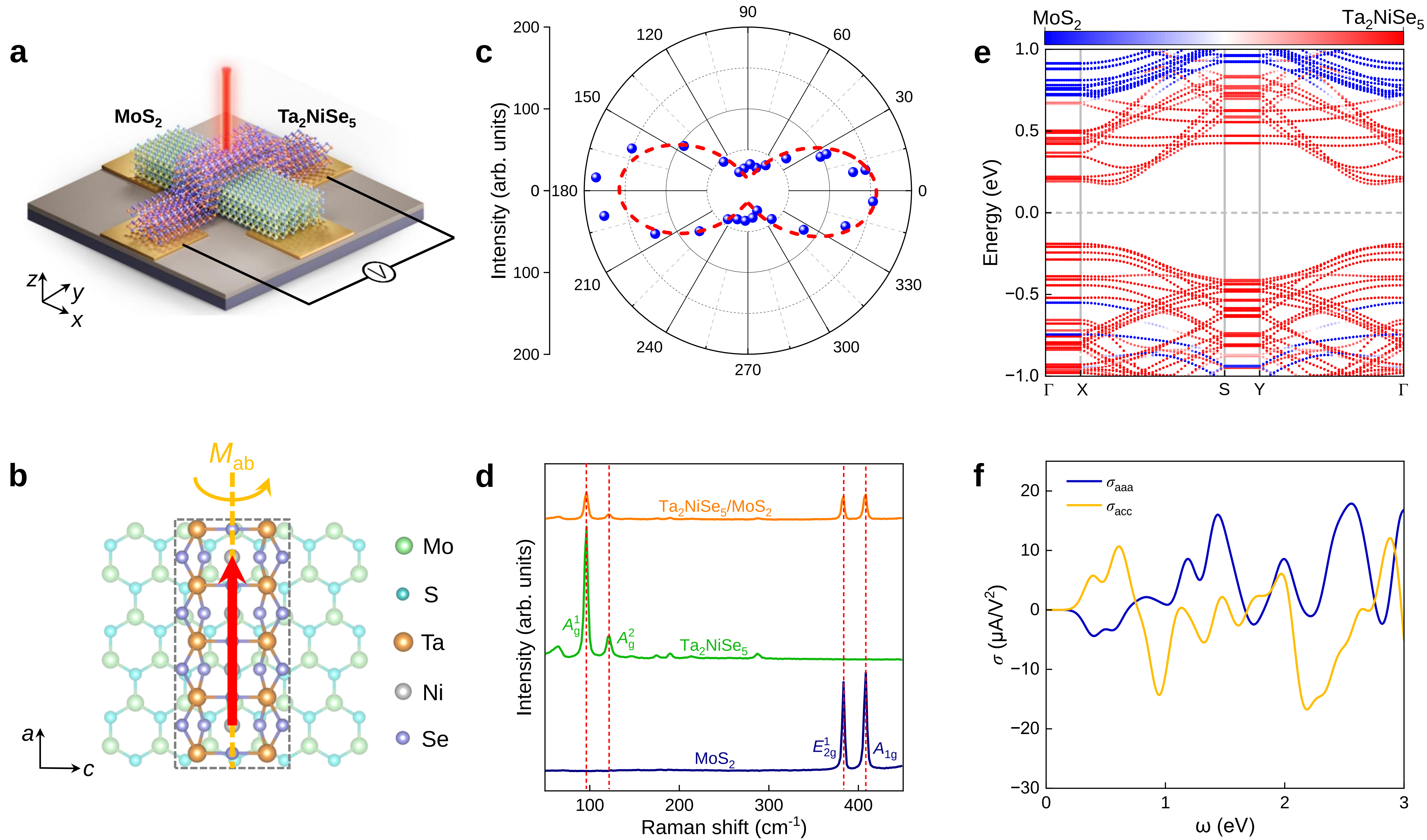
a
MoS2
Ta2NiSe5
V
z
y
x
b
Mab
Mo
S
Ta
Ni
Se
a
c
c
Intensity (arb. units)
90
120
60
150
30
180
0
210
330
240
300
270
200
100
0
100
200
d
Ta2NiSe5/MoS2
Ta2NiSe5
MoS2
Intensity (arb. units)
Raman shift (cm-1)
100
200
300
400
e
MoS2
Ta2NiSe5
Energy (eV)
1.0
0.5
0.0
−0.5
−1.0
Γ
X
S
Y
Γ
f
σaaa
σacc
σ (μA/V2)
20
10
0
−10
−20
−30
ω (eV)
0
1
2
3

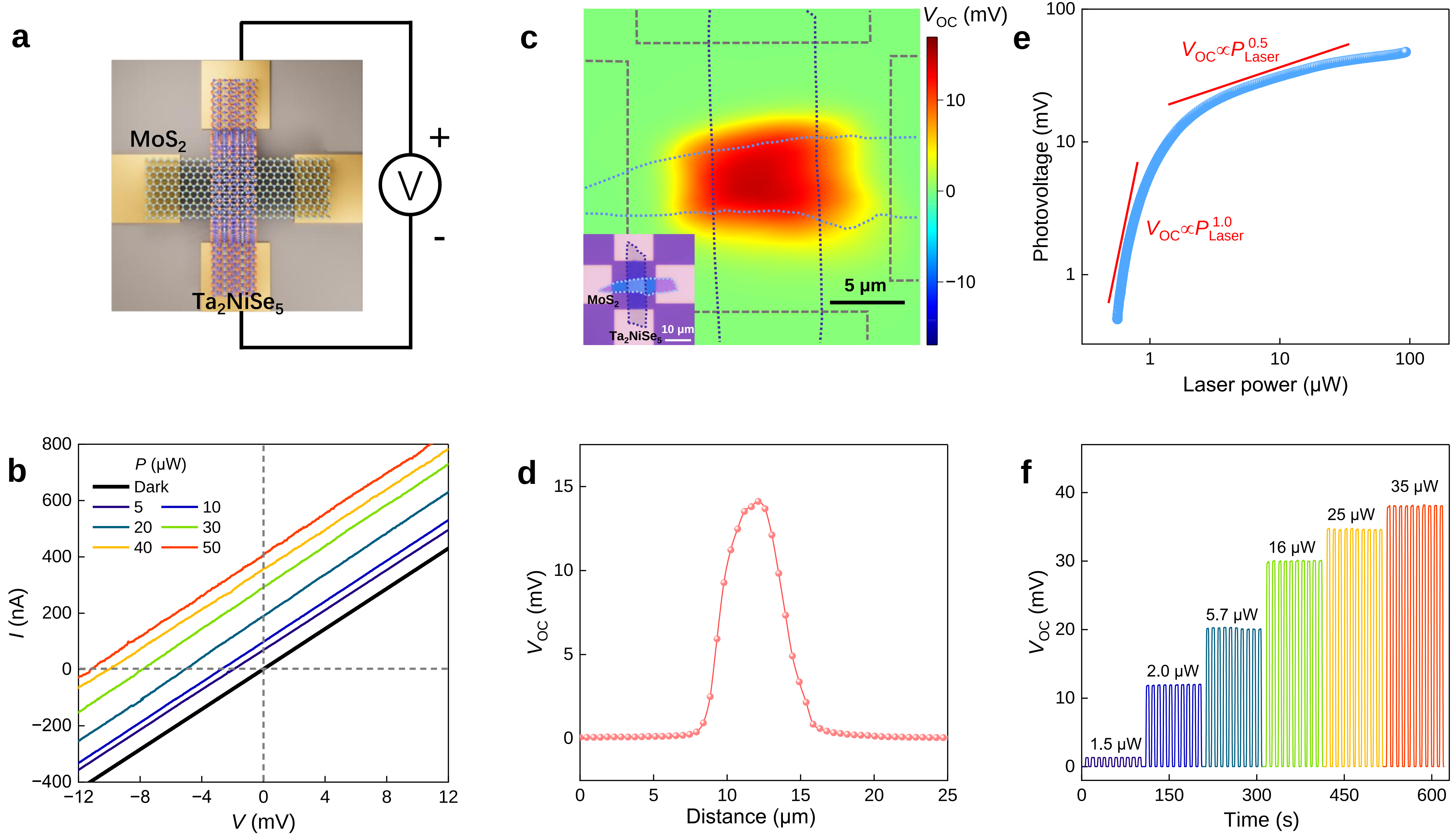
a
MoS2
Ta2NiSe5
V
+
-
b
P (μW)
Dark
5
10
20
30
40
50
I (nA)
V (mV)
c
VOC (mV)
10
0
−10
5 μm
MoS2
Ta2NiSe5
10 μm
d
VOC (mV)
Distance (μm)
e
VOC∝P 0.5 Laser
VOC∝P 1.0 Laser
Photovoltage (mV)
Laser power (μW)
f
1.5 μW
2.0 μW
5.7 μW
16 μW
25 μW
35 μW
VOC (mV)
Time (s)

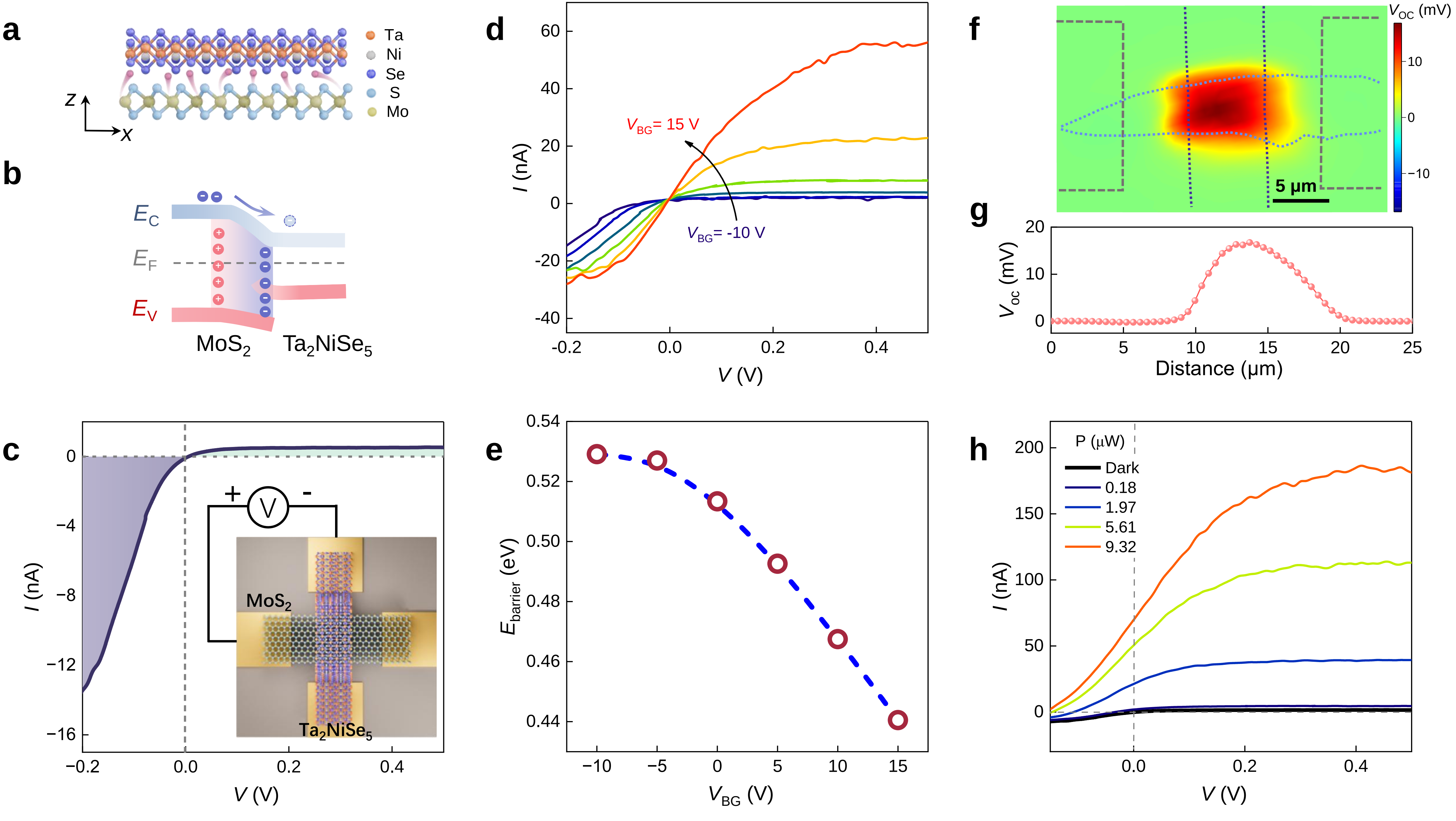

a
Ta
Ni
Se
S
Mo
z
x
b
$E_C$
$E_F$
$E_V$
$MoS_2$
$Ta_2NiSe_5$
c
$I$ (nA)
$V$ (V)
+
V
-
$MoS_2$
$Ta_2NiSe_5$
d
$V_{BG}$= 15 V
$V_{BG}$= -10 V
$I$ (nA)
$V$ (V)
e
$E_{barrier}$ (eV)
$V_{BG}$ (V)
f
$V_{OC}$ (mV)
5 μm
g
$V_{oc}$ (mV)
Distance (μm)
h
P (μW)
Dark
0.18
1.97
5.61
9.32
$I$ (nA)
$V$ (V)

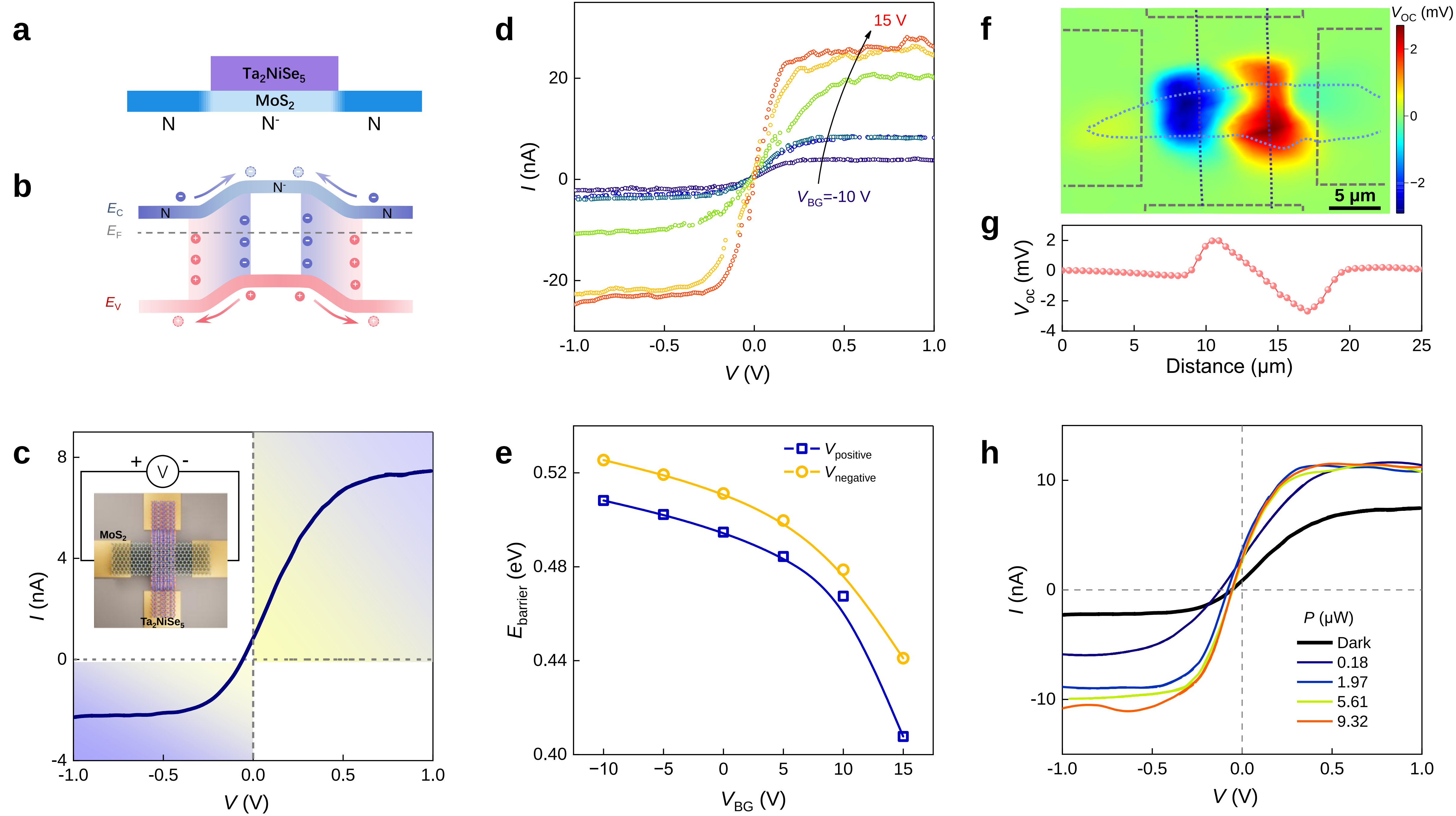

a
Ta2NiSe5
MoS2
N
N-
N
b
EC
EF
EV
N-
N
N
c
8
4
0
-4
I (nA)
V
+
-
MoS2
Ta2NiSe5
-1.0
-0.5
0.0
0.5
1.0
V (V)
d
15 V
VBG=-10 V
20
0
-20
I (nA)
-1.0
-0.5
0.0
0.5
1.0
V (V)
e
Vpositive
Vnegative
0.52
0.48
0.44
0.40
Ebarrier (eV)
−10
−5
0
5
10
15
VBG (V)
f
VOC (mV)
2
0
−2
5 μm
g
2
0
-2
-4
Voc (mV)
0
5
10
15
20
25
Distance (μm)
h
10
0
-10
I (nA)
P (μW)
Dark
0.18
1.97
5.61
9.32
-1.0
-0.5
0.0
0.5
1.0
V (V)

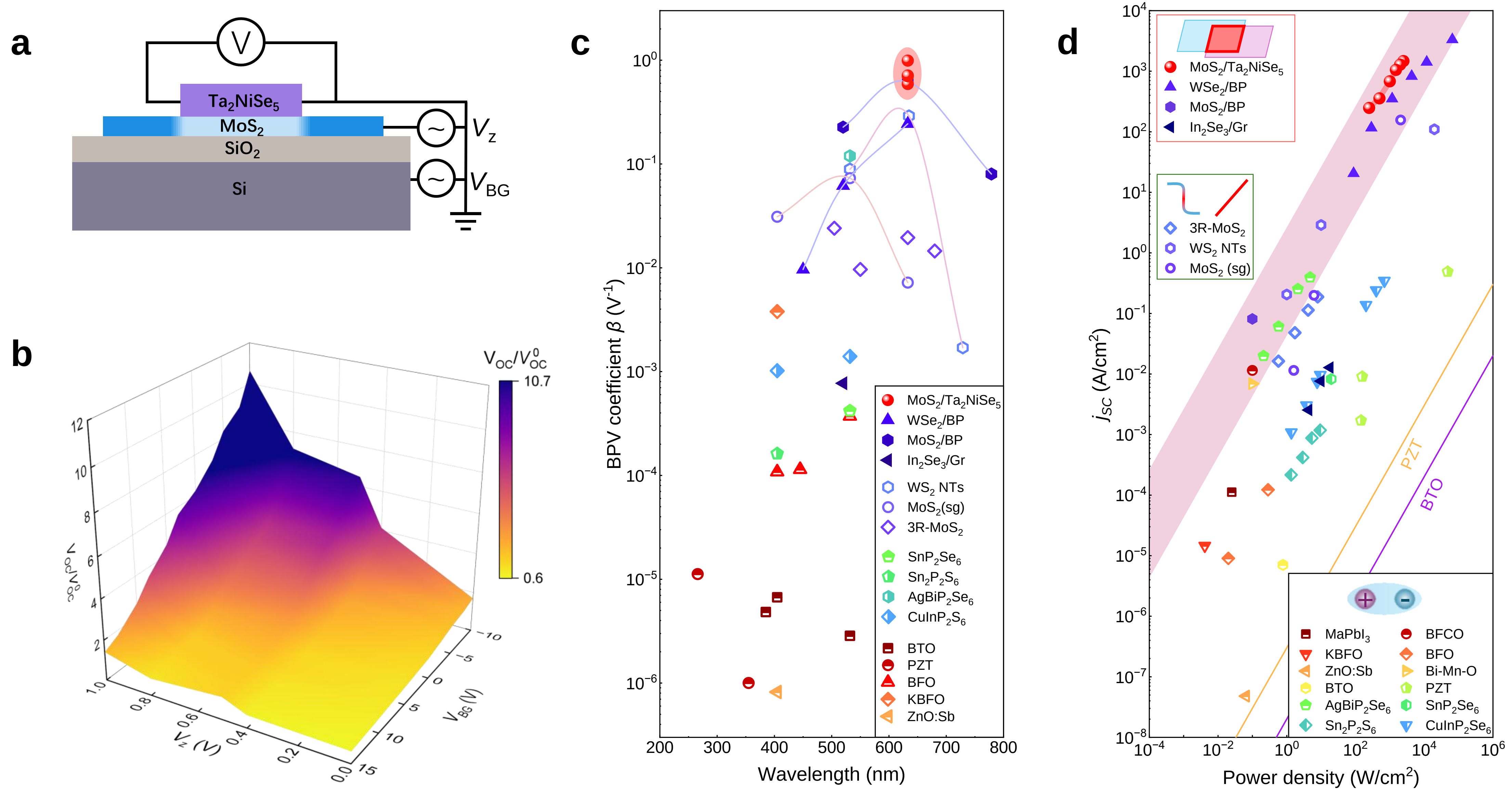

a
V
Ta2NiSe5
MoS2
SiO2
Si
Vz
VBG
b
VOC/V0OC
10.7
0.6
VOC/V0OC
Vz (V)
VBG (V)
c
BPV coefficient β (V-1)
Wavelength (nm)
MoS2/Ta2NiSe5
WSe2/BP
MoS2/BP
In2Se3/Gr
WS2 NTs
MoS2(sg)
3R-MoS2
SnP2Se6
Sn2P2S6
AgBiP2Se6
CuInP2S6
BTO
PZT
BFO
KBFO
ZnO:Sb
d
jSC (A/cm2)
Power density (W/cm2)
MoS2/Ta2NiSe5
WSe2/BP
MoS2/BP
In2Se3/Gr
3R-MoS2
WS2 NTs
MoS2 (sg)
PZT
BTO
MaPbI3
BFCO
KBFO
BFO
ZnO:Sb
Bi-Mn-O
BTO
PZT
AgBiP2Se6
SnP2Se6
Sn2P2S6
CuInP2Se6